\documentclass[aps,prb,twocolumn,superscriptaddress]{revtex4}
\usepackage{graphics}
\usepackage{longtable}
\usepackage[epsfig]{graphicx}
\usepackage{epsfig}
\usepackage{amsmath}
\usepackage{float}
\begin{document}
	
	\title{Superconductivity in intercalated buckled two-dimensional materials: KGe$_2$}
	
	\author{Sherif Abdulkader Tawfik}
	\email{sherif.abbas@uts.edu.au}
	\affiliation{School of Mathematical and Physical Sciences, University of Technology Sydney, Ultimo, New South Wales 2007, Australia.}
	\affiliation{Institute for Biomedical Materials and Devices (IBMD), Faculty of Science, University of Technology Sydney, Sydney, NSW, Australia.}

	\author{Catherine Stampfl}
	\affiliation{School of Physics, The University of Sydney, Sydney, New South Wales, 2006, Australia.}

	\author{Michael J. Ford}
	\email{mike.ford@uts.edu.au}
	\affiliation{School of Mathematical and Physical Sciences, University of Technology Sydney, Ultimo, New South Wales 2007, Australia.}
	\begin{abstract}
		Germanene has emerged as a novel two-dimensional material with various interesting properties and applications. Here we report the possibility of superconductivity in a stable potassium intercalated germanene compound, KGe$_2$, with a transition temperature $T_c \sim 11$ K, and an electron-phonon coupling of 1.9. Applying a 5\% tensile strain, which reduces the buckling height by 4.5\%, leads to the reduction of the electron-phonon coupling by 11\% and a slight increase in $T_c \sim 12$ K. That is, strong electron-phonon coupling results from the buckled structure of the germanene layers. Despite being an intercalated van der Waals material similar to intercalated graphite superconductors, it does not possess an occupied interlayer state.
		
	\end{abstract}
	
	\maketitle
\section{Introduction} 
The largest class of high-temperature superconductors includes materials that are formed of intercalated two-dimensional (2D) structures, such as YBa$_2$Cu$_3$O$_7$ and iron-based superconductors. The simplest such structure is MgB$_{2}$, in which a layered flat hexagonal material is intercalated by a single element, which was discovered in 2001\cite{MgB2Review} to have a high $T_{c}$ of 39 K. In 2005, another class of intercalated 2D materials was found to superconduct: graphite intercalated with Ca and Yb.\cite{10} This class of structures, known as graphite intercalated compounds (GIC), is fundamentally different to MgB$_{2}$, because it is formed of graphene layers that can exist individually as a stable crystal. Since 2005, the number of new GICs has not increased; to date, the only known superconducting GICs are those with intercalatants K, Ca, Li, Yb, Sr and Ba. This discovery created immense research interest aiming to understand the mechanism that underlies the superconductivity in this class of materials. 

In spite of the structural similarity of GICs to MgB$_{2}$, the superconducting temperature of MgB$_{2}$ is far higher than all of the observed $T_c$'s of GICs, the maximum on record being 11.5 K for CaC$_6$.\cite{10} Moreover, it was observed that a superconducting gap appears on the Fermi surface associated with the intercalatant atom, but not in the $\pi^{*}$ orbital of the graphitic layers,\cite{key-2} confirming the theoretical prediction that there is an occupied interlayer state in all superconducting compounds of this class of materials.\cite{key-3} While MgB$_2$ possesses an interlayer state, it is unoccupied. Another key difference is the fact that MgB$_2$ is a two-gap superconductor,\cite{MgB2Review} and the two gaps were theoretically shown\cite{EPW} to arise from the $\sigma$ and $\pi$ bands. However, graphite intercalated compounds are single gap superconductors.

Nevertheless, graphite intercalation compounds and MgB$_2$ have a number of features in common. The hexagonal 2D layers in both are perfectly flat; that is, there is no buckling. In fact, buckling was thought to destroy the superconducting state in layered Li$_x$BC because it induces strong mixing of the $\sigma$ and $\pi$ bands.\cite{LixBC} The calculated electron-phonon coupling (EPC) strength, $\lambda$, of bulk CaC$_6$ is 0.83,\cite{key-4} and that of MgB$_2$ is 0.748 \cite{EPW}, so the EPC strength in both compounds are quite close. 

It is interesting to consider what would then be the superconducting properties of an intercalated vdW material that is intrinsically buckled? An example of a buckled 2D material is germanene, which can be viewed as the cleaved (111) layer of the $Fd3m$ phase of bulk germanium. It is predicted to be a stable 2D Dirac material.\cite{Germanene} Here we explore the potential superconductivity of potassium-intercalated germanene, KGe$_2$, which is a hypothetical compound that resembles the structure of CaGe$_2$ that is already well known.\cite{CaGe2_1} The interesting feature of KGe$_2$ is that the germanene layers preserve their Dirac cones; that is, KGe$_2$ is a truly intercalated 2D germanene material. The K intercalation was also observed to enhance the superconductivity of FeSe,\cite{key-7} and the K intercalation of MoS$_2$ leads to the emergence of several superconducting phases.\cite{4} We find by solving the anisotropic Eliashberg equations based on density-functional theory (DFT) and time-dependent perturbation theory\cite{EPW} that the superconducting gap is $\sim 11$ K, which is close to that observed in CaC$_6$.

%\section{Computational details}
\section{Computational details}
 The DFT calculations are performed using the local density approximation (LDA)\cite{LDA} and norm-conserving pseudopotentials\cite{TM} using QUANTUMESPRESSO\cite{espresso}. The valence electronic wave functions are expanded in a plane-wave basis set with a kinetic energy cutoff of 40 Ry. We use a $12\times 12\times 12$ \textbf{k}-point mesh for KGe$_2$ and $12\times 12\times 1$ for monolayer germanene, and a Methfessel-Paxton smearing\cite{MP} of 0.10 eV. The dynamical matrices and the linear variation of the self-consistent potential are calculated within density-functional perturbation theory\cite{LR} on the irreducible set of a regular $12\times 12\times 12$ \textbf{k}-point mesh and a $4\times 4\times 4$ \textbf{q}-point mesh for KGe$_2$, and  $12\times 12\times 1$ \textbf{k}-point mesh and a $4\times 4\times 1$ \textbf{q}-point mesh for germanene. In order to solve the Eliashberg equations we evaluate electron energies, phonon frequencies, and electron-phonon matrix elements on fine grids with a $N_k=20\times 20 \times 20$ Monkhorst-Pack mesh and a $N_q=20\times 20 \times 20$ for KGe$_2$ and $N_k=20\times 20 \times 1$ Monkhorst-Pack mesh and a $N_q=20\times 20 \times 1$ for germanene, which were obtained by convergence of the EPC. The calculations are performed using smearing parameters in the Dirac $\delta$ functions corresponding to 100 and 0.5 meV for electrons and phonons, respectively.

\def\d{{\bf d}}
\def\k{{\bf k}}
\def\q{{\bf q}}
\def\G{{\bf G}}
\def\R{{\bf R}}

The EPC, $\lambda$, is calculated according to the equation
\begin{equation}
\lambda = 2 \int {\alpha^2F(\omega) \over \omega} d\omega.
\end{equation}

\noindent where the Eliashberg spectral function $\alpha^2F(\omega)$ is defined as

\begin{equation}
\alpha^2F(\omega) = {1\over 2\pi N(e_F)}\sum_{\q\nu} 
\delta(\omega-\omega_{\q\nu})
{\gamma_{\q\nu}\over\hbar\omega_{\q\nu}}\quad,
\end{equation}

\noindent where $\omega$ is the phonon frequency, $\q$ and $\nu$ are the phonon momentum and mode, respectively, $N(e_F)$ is the electronic density of states at the Fermi level, and $\gamma_{\q\nu}$ the electron-phonon coupling strength associated with
a specific phonon mode $\nu$ and momentum $\q$. 

%In order to quantify the different contributions to the coupling strength associated with the ? and ? sheets of the Fermi surface we evaluate a band- and wave-vector-dependent electron-phonon coupling strength defined by

%\section{Results and discussions}

%Geometry:
%A word on Ge2: 

%Are there known polymorphs? List the computational ones.
%So far there have been known K-Ge polymorphs in existance. However, the computational database, such as MaterialsProject and AFLOW, host a number of potential polymorphs. The cohesive energies of these polymorphs are .

\section{Results and discussions} 

\begin{figure}[h]
	\includegraphics[width=80mm]{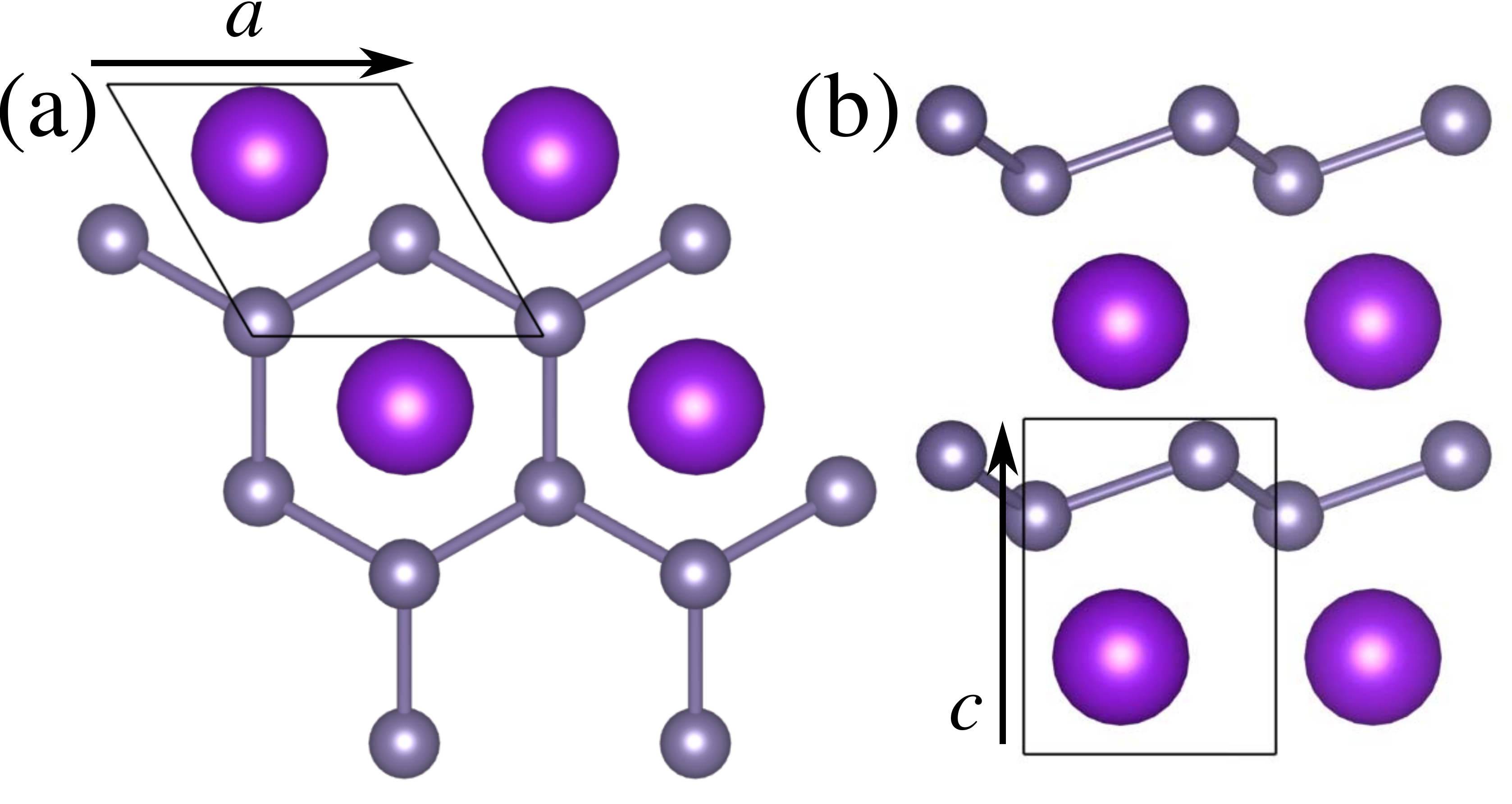}
	\caption{The (a) top and (b) side view of the KGe$_2$ structure with the $AA$ stacking. The unit cell is indicated with associated lattice vectors. The K atom is positioned in the center of the germanene hole.}
	\label{fig_Structure}
\end{figure}

In all our calculations of the layered KGe$_2$ we have used the $AA$ stacking of the germanene layers as displayed in Fig. \ref{fig_Structure}. The $AA$ stacking was found to be more favorable than the $AB$ stacking based on comparing the stabilization energies of the two configurations, $E_{S}$, which is defined as $E_{S}=E_{KGe_2}-E_{Ge_2}-E_{K}$, where $E_{KGe_2}$ is the total energy of KGe$_2$, $E_{Ge_2}$ is the total energy of the germanene unit cell and $E_{K}$ is the total energy of an isolated K atom. The $E_{S}$ of $AA$ was calculated to be -2.46 eV, which is 126 meV lower than that of $AB$. Finding the optimal stacking sequence for intercalated layered compounds is a daunting task, we choose here to only study the $AA$ stacking based on the following considerations: This unit cell is the smallest of all KGe$_2$ structures, which offers computational efficiency. Even though other stacking sequences are possible, the $AA$ stacking is a good representative model of the layered KGe$_2$ for the purpose of calculating the superconducting properties, because the small horizontal shifts in the position of the K atoms across the layers are expected to have little effect on its superconducting properties \cite{27}. The calculated lattice parameters of the $AA$ KGe$_2$ structure are $a=3.987$ \AA and $c= 4.596$ \AA, while that of monolayer germanene is $a=3.921$ \AA. The buckling height of the germanene layers in KGe$_2$ (the $z$-axis distance between the two Ge atoms) is 0.827 \AA, while that of monolayer germanene is 0.620 \AA, in agreement with published results.\cite{Germanene} With such a crystal, the K-K bond length becomes 3.987 \AA, which is higher than the bond length of the K-K bond (3.577 \AA) in the \textit{Im3m} K polymorph at 12 GPa.\cite{K12GPa}

In order to examine the impact of buckling on the EPC and the superconducting properties, we applied a planar strain on KGe$_2$. Specifically, we tested the strains along the $a$ axis of $\pm 5$\% and $\pm 10$\%. We found that $- 5$\% and $- 10$\% tensile strains result in dynamically unstable structures (that is, the phonon dispersion has imaginary frequencies). In the case of positive strains, a 5\% strain (with the lattice parameter $a=4.212$ \AA) maintains the dynamical stability of the structure, whereas 10\% strain yields a dynamically unstable structure. Therefore, we focus here on the structure with 5\% positive strain. The buckling height in this structure is 0.790 \AA, which is 4.5\% less than that of the unstrained structure.

%To find the lattice constant for this structure, we applied three different DFT methods: LDA, PBE, and PBE-TS. Tab. I displays the results. It is clear that the vdW force plays an important role in this structure, given that the PBE method yields a much larger $c$ lattice constant than the PBE-TS method. Given that the LDA gives a geometry that is closet oth PBE-TS than PBE geometry, we use the LDA method in the subsequent calculations.

%\subsection{Electronic properties}

\begin{figure}[h]
	\includegraphics[width=80mm]{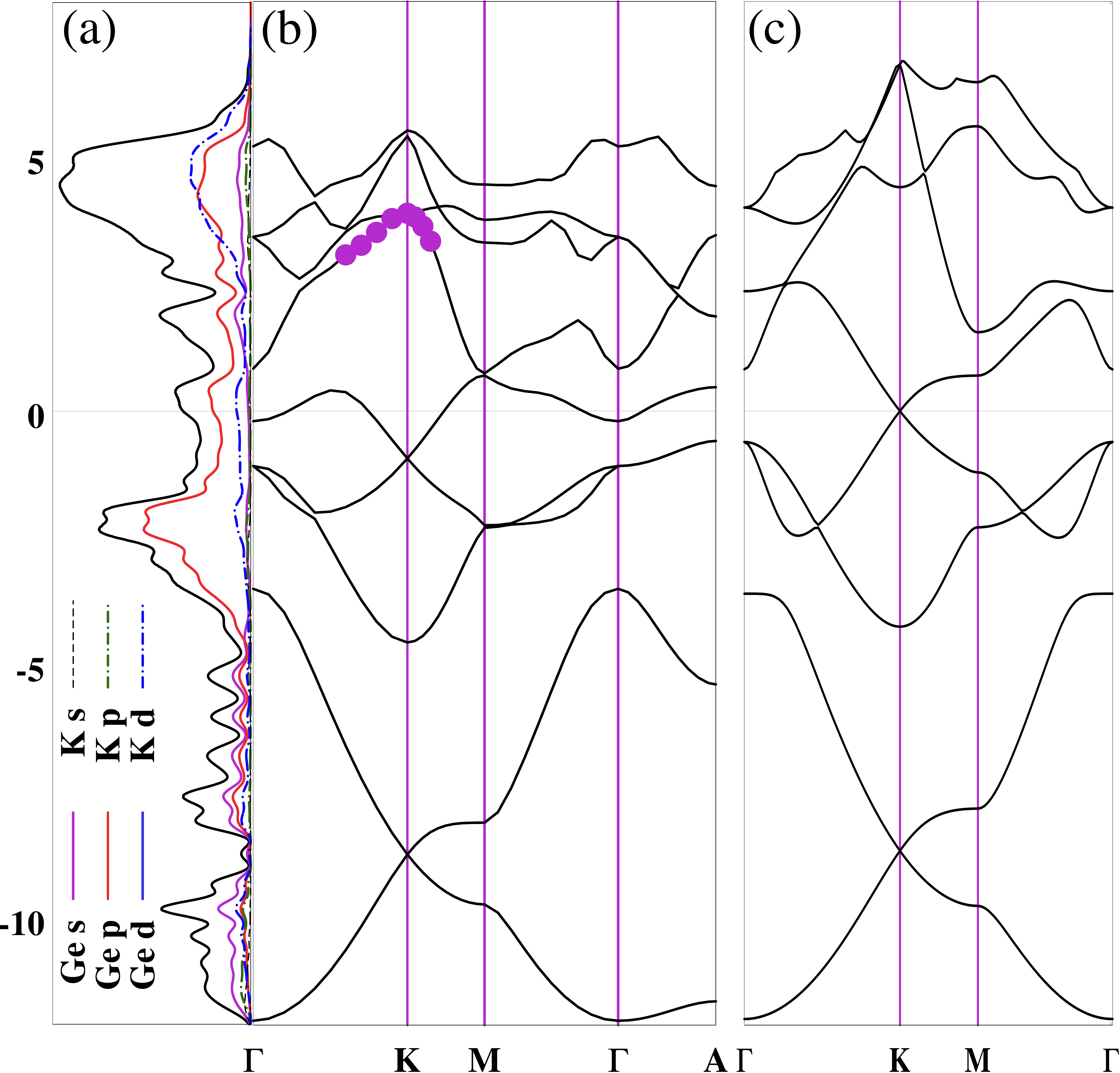}
	\caption{(a) The partial electronic density of states (PDOS) of KGe$_2$, (b) the KGe$_2$ band structure, indicating the orbitals with an interlayer nature as filled circles, and (c) the band structure of monolayer germanene. The three plots are aligned at the Fermi level, which is the energy zero.}
	\label{fig_bands}
\end{figure}

With the buckling of the germanene layers in KGe$_2$, the coupling between the $\pi$ band and the K band is expected to be much stronger than in MgB$_2$ and CaC$_6$. The band structure of KGe$_2$ and the projected density of states are presented in Fig. \ref{fig_bands}(a,b), and the band structure of monolayer germanene is diplayed in Fig. \ref{fig_bands}(c). In the unit cell, according to Bader charge analysis, the K atom donates 1 $\lvert e\lvert$ to the germanene layer, which results in shifting the Dirac point of germanene in Fig. \ref{fig_bands}(b) downwards. This full charge transfer is analogous to the nearly-full charge transfer that takes place in KC$_8$ \cite{KC8}.

As the Dirac cone is shifted below the Fermi level, there is one Ge $4p$ band, that is slightly hybridized with K $3p$, crossing the Fermi energy. Compared to the CaC$_6$ band structure, the Dirac cone in KGe$_2$ stays intact, while in CaC$_6$ the $\Gamma$ point is opened by $\sim 0.5$ eV. The Dirac cones here do not experience a momentum shift, unlike the situation in CaSi$_2$ (which is non-superconducting) where the Dirac cone shifts slightly from the high symmetry points of the first Brillouin zone due to the electron transfer between the adjacent silicene layers \cite{CaSi2}. The situtation of KGe$_2$ is akin to KC$_8$ \cite{KC8} (a superconductor), and electrostatically doped graphene, where only minor differences related to the intercalation states are present close to Fermi energy at the $\Gamma$ point.
%\subsection{The definition of an interlayer state}

It is of interest to determine whether KGe$_2$ exhibits an interlayer state like intercalated graphite superconductors? The presence of such a state in superconducting GIC was first realized as a striking ``coincidence'' in Ref. \citenum{key-3} and is characterized as a hybridized band with significant charge density in the interlayer region. In the case of KGe$_2$, as shown by the small filled circles in Fig. \ref{fig_bands}(b), there is no occupied band that is dominated by an interlayer nature. Instead, the occupied band in Fig. \ref{fig_bands}(b) only has a few momentum points that have an interlayer character as determined by inspecting the wave functions.

%We examine the possiblity of the existence of an antiferromagnetic (AFM) state by enforcing the magnetic moment directions on the Ge atoms. We arrived at a AFM ground state for KGe$_2$ in which the two Ge atoms are opposite in direction, each carrying a spin of 0.12 $\mu_B$, and the difference in energy between the AFM and the non-spin polarized configuration is only 72 meV. The AFM state is, however, not permitted in CaC6 or MgB2, but a residual spin polarization appears in KGe$_2$ as a result of the buckling, where the Ge $4p$ orbitals 

%\subsection{Vibrational and superconducting properties}

%The primitive cell has $C_{3}$ symmetry and contains 3 atoms, leading to 9 vibrational modes. 

%Except for acoustic modes, a clear separation exists between optical Li and B vibrations. Li modes are confined in the 40-55 meV region and are not dispersive, meaning that Li-vibrations behave essentially as Einstein modes. Boron in-plane vibrations are softened along the T direction due to coupling to the bands. The softening at of the Eg phonon branches is approximately 20 meV, to be compared with the 25-30 meV in MgB2 for the E2g modes.5,37,38 This suggests a strong coupling of the bands to the in-plane vibrations in LiB,25 almost as strong or comparable to that of MgB2.

\begin{figure}[h]
	\includegraphics[width=80mm]{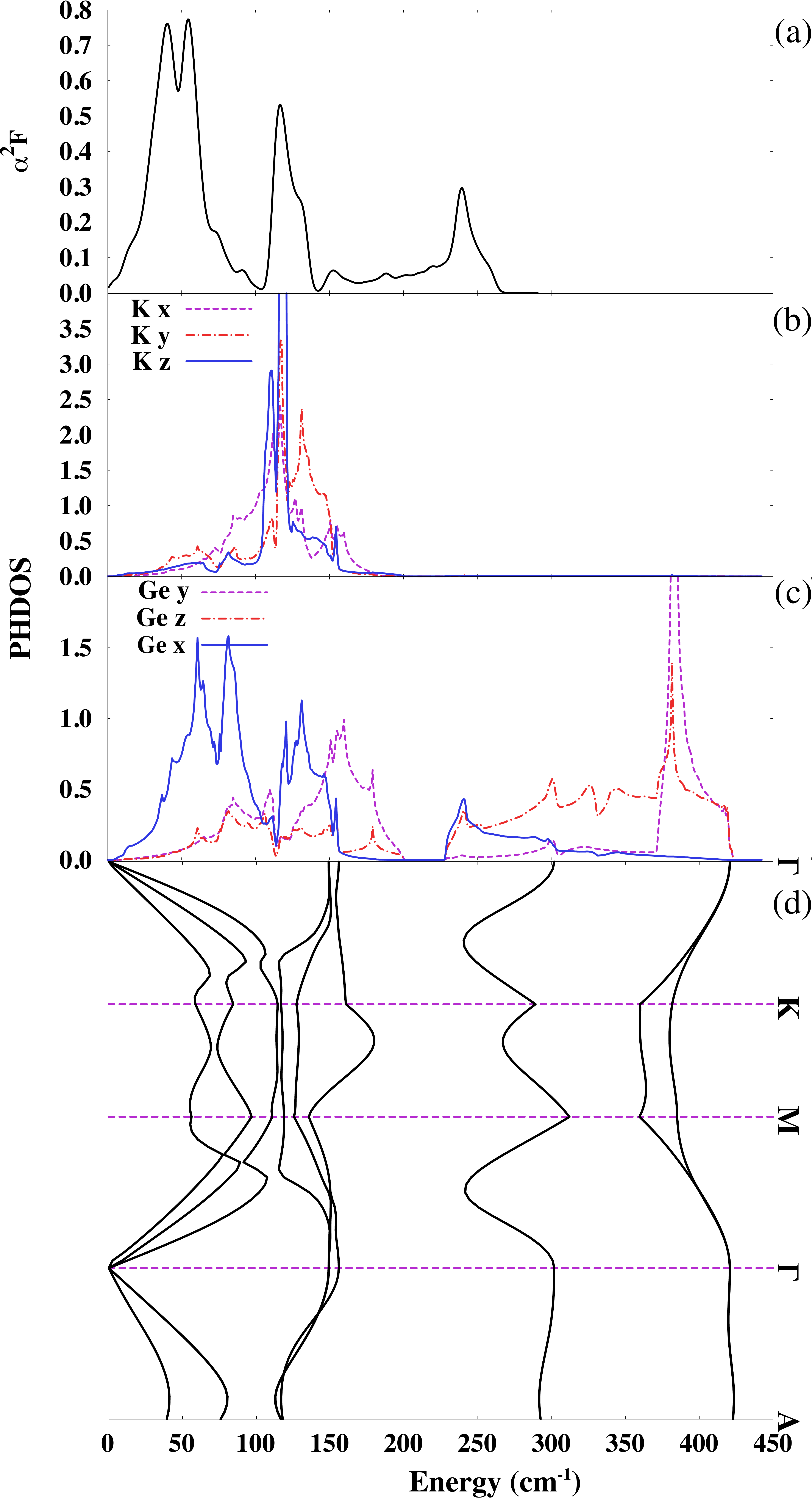}
	\caption{(a) The Eliashberg function $\alpha^2 F(\omega)$ of KGe$_2$, the atom-projected phonon density of states (PHDOS) for the (b) K and the (c) Ge atoms, and the (d) phonon dispersion along the high symmetry points of the Brillouin zone.}
	\label{fig_ProjPHDOS}
\end{figure}

Figure \ref{fig_ProjPHDOS}(a) displays the isotropic Eliashberg spectral function $\alpha^2 F(\omega)$ for KGe$_2$. $\alpha^2 F(\omega)$ displays a large dominant peak centered around 50 meV, a second weaker peak centered around 120 meV, and a third weaker peak centered at 240 meV. The corresponding isotropic electron-phonon coupling strength is $\gamma = 1.90$. In order to undestand the vibrational origin of these peaks, we display the atom-projected phonon density of states (PHDOS) in Fig. \ref{fig_ProjPHDOS}(b,c). The $\alpha^2 F(\omega)$ peak centered at 50 meV originates from Ge vibrations, mainly from the Ge out-of-plane modes. The peak centered at 120 meV originate from K and Ge modes. Regarding K, the main contributing mode is the out-of-plane component, then the planar modes in both the $x$ and $y$ directions. For the Ge contributing modes, again the modes in all directions contribute, where the out-of-plane modes contribute more than the planar modes. Finally, the peak centered at 240 meV does not have any K contribution. It is driven by Ge $z$ and $y$ modes. That third peak does not have an influence on the value of $\lambda$, as can be seen in the flattening of the $\alpha^2 F(\omega)$ curve beyond 200 meV.

The $\alpha^2 F(\omega)$ of KGe$_2$ is different to that of CaC$_6$,\cite{key-4} where the $\alpha^2 F(\omega)$ of the latter has three primary peaks, the low energy peak being contributed mainly by Ca planar modes, while the second peak is contributed by C out-of-plane modes, and the third C peak is contributed by planar modes. First of all, the planar modes in both directions contribute equally in CaC$_6$ due to the lattice symmetry, unlike the case of KGe$_2$ where the positions of the two Ge atoms with respect to the K atom within unit cell are not symmetric. Second, given that the K atom is lighter than the Ge atom, the first Ge peak has a lower energy than the K peak, which is opposite to the case of CaC$_6$, where the first C peak has a higher energy than the Ca peak. Third, the lowest-energy peak in CaC$_6$ is almost purely Ca dominated, mostly of planar modes, whereas the K peak has a mixture of K and Ge modes, the majority of which are out-of-plane modes.

\begin{figure}[h]
	\includegraphics[width=80mm]{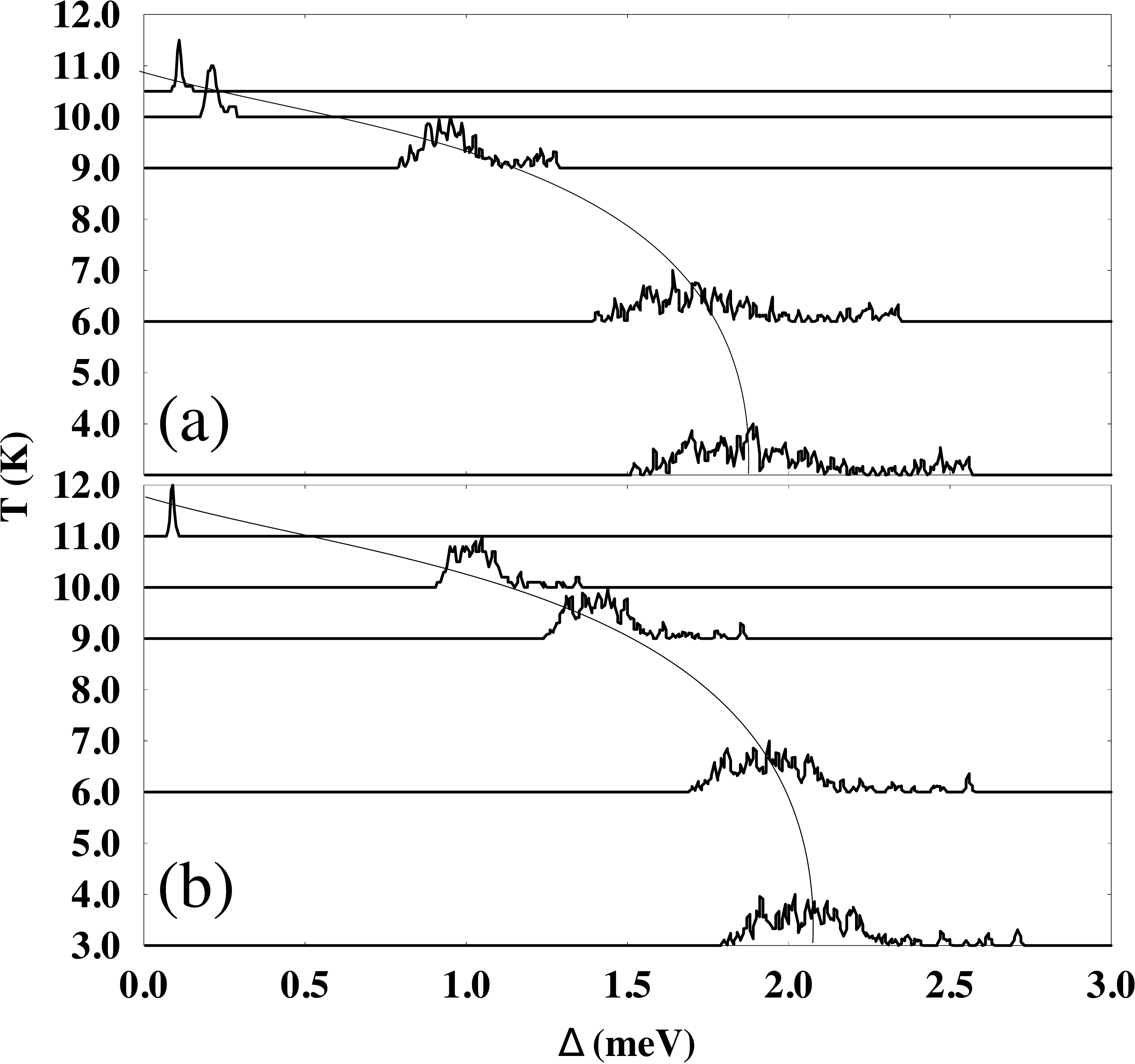}
	\caption{The superconducting gap function $\Delta(\omega)$ at various values for $T$ for the (a) equilibrium and (b) 5\% tensile-strained KGe$_2$. The curved lines show the trend as the values of $\Delta(\omega)$ converge towards zero.}
	\label{fig_Delta}
\end{figure}

The calculated electron-phonon coupling is much larger than the values reported for MgB$_2$ and the intercalated graphite compounds, but is closer to the range of values of strong-coupling superconductors such as Pd \cite{EPW}. We display a list of these values in Tab. I for 2D intercalated compounds. The reason for this difference is the large buckling of the germanene layer, which leads to the enhanced EPC. The situation in KGe$_2$ is in stark contrast to the CaGe$_2$ compound, where the latter has a very small EPC of 0.19 (cf. Tab I) and does not superconduct.

Within the anisotropic Eliashberg formalism \cite{EPW}, the value of $T_c$ is obtained by examining the gap function $\Delta$ as the temperature parameter $T$ is changed. When $\Delta$ vanishes at some $T$, this characterizes the superconducting state, at which $T_c=T$. Such identification is performed by plotting the $\Delta(\omega)$ function at various values of $T$, and inspecting the trend of the function as its peaks converge to zero, as discussed in \cite{EPW} and displayed in Fig. \ref{fig_Delta}. This figure that the $\Delta(\omega)$ converges to zero as $T$ appraches $T_c \sim 11$ K. The convergence trend is displayed by a straight line that passes through the various $\Delta(\omega)$ functions.

In the isotropic limit, the Eliashberg formalism reduces to the Allen-Dynes formulation \cite{Allen}, in which $T_c$ is given by 

\begin{equation}
T_c = {\omega_{log}\over 1.2} \mbox{exp} \left [ 
{-1.04(1+\lambda)\over \lambda(1-0.62\mu^*)-\mu^*}\right ]
\label{allendynes}
\end{equation}

\noindent where $\mu^*$ is the Coulomb pseudopotential, for which we use $\mu^*=0.16$, and the phonon frequencies logarithmic average $\omega_{log}$ is given by

\begin{equation}
\omega_{log} = \mbox{exp} \left [ {2\over\lambda} \int {d\omega\over\omega}
\alpha^2F(\omega) \mbox{log}\omega \right ]
\end{equation}

\noindent The value of $T_c$ calculated for KGe$_2$ using the isotropic Allen-Dynes formalism (Eq. \ref{allendynes}) is 5.8 K, which is almost half of the value predicted by solving the full Eliashberg equation. This is because of the significance of the momentum anisotropy in KGe$_2$, which is also the case in MgB$_2$.

\begin{table}
	\label{tab:comparison}
	\caption{The electron-phonon coupling strength $\lambda$, the predicted and experimental superconducting critical temperatures $T_c$ for a number of 2D-intercalated compounds.}
	%\begin{ruledtabular}
	\begin{tabular}{|c|c|c|c|c|c|c|}
		\hline

Structure	&		$\lambda$	&	$T_c$	&	$T_c$ exp	& Ref.	\\
		\hline
KGe$_2$	&		1.9	&	11	&		&	Present work \\
MgB$_2$	&		0.748	&		50	&	39	&	\citenum{EPW} \\
CaC$_6$	&		0.83	&		11	&	11.5	&\citenum{key-4} \\ 
CaC$_2$ [95 GPa]		&	0.564	&	9.8	&		&	\citenum{CaC2}\\
LiB	&	0.62	&	10-15	&		&	\citenum{LiB} \\ 
%CaGe$_2$ & 0.19 & & &	\citenum{SherifYiDu} \\ 
		\hline
	\end{tabular}
	%\end{ruledtabular}
\end{table}

Another model for the superconductivity of intercalated compounds is that proposed by Al-Jishi \cite{Jishi} for GICs. This is a simplified BCS-based purely electronic model where the graphite $\pi$ and intercalatant $s$ states of are coupled via a coupling parameter. We can easily extend this model to KGe$_2$, where the Hamiltonian couples between the germanene $\pi$ and the K $4s$ states, and we obtain the equation

\begin{equation}
kT_c \sim \hbar \omega_c exp(-\frac{1}{\arrowvert \lambda \arrowvert}\sqrt{N_{\pi}(0)N_{4s}(0)}),
\label{jishi}
\end{equation}

\noindent where $\omega_c$ is the Debye frequency, $N_{\pi}(0)$ and $N_{4s}(0)$ are the density of states of the $\pi$ and the K $4s$ states at the Fermi level, respectively. The critical feature in this equation is that $T_c$ would become zero when one of $N_{\pi}(0)$ and $N_{4s}(0)$ is zero. In our KGe$_2$, full charge transfer occurs from the K atom to the germanene layer, which should lead to a $T_c\sim 0$. The reason is that, with increasing charge gain in the graphite layers in GICs, the 2D electrons screen the polar coupling between the intercalatant atoms \cite{Takada}. This, however, is not the case in KGe$_2$, owing to the large electron-phonon coupling contribution arising from the electron-doped germanene layers; that is, superconductivity here is $\pi$-driven (most of $\lambda$ is contributed by the germanene layers), in contrast to the case of GICs where superconductivity is interlayer-driven (most of $\lambda$ is contributed by the Ca atoms) \cite{key-4}.

%An alternative simplified model, proposed by Takada \ref{Takada} for GICs, is based on representing the electron-phonon interaction by the oscillations of the positively charged intercalatant and negatively charged C ions. In this model, $T_c$ is calculated by solving the gap equation

\section{Conclusions} 

We predict a superconducting temperature of $\sim 11$ K in a novel buckled intercalated compound, KGe$_2$. The compound has a large electron-phonon coupling of 1.9, which decreases by 11\% when a positive planar tensile strain of 5\% is applied. This is acompanied by a slight increase in $T_c$ of $\sim 12$ K. That is, strong electron-phonon coupling results from the buckled structure of the germanene layers. Despite being an intercalated van der Waals material like intercalated graphite superconductors, KGe$_2$ does not possess an occupied interlayer state.

This research was funded by the Australian Government through the
Australian Research Council (ARC DP160101301). Theoretical
calculations were undertaken with resources provided by the National
Computational Infrastructure (NCI) supported by the Australian
Government and by the Pawsey Supercomputing Centre funded by the
Australian Government and the Government of Western Australia.

\end{document}